\documentclass[aps,pre,twocolumn,array,epsfig,eqsecnum]{revtex4}
\usepackage{amsmath}
\usepackage{amsfonts}
\usepackage[dvips]{graphicx}

\begin{document}   
\setlength{\parindent}{0pt}

\title{Relevance of rank for a mixed state quantum teleportation resource}

\author{K.G. Paulson and S.V.M. Satyanarayana}
\address{
Department of Physics, Pondicherry University, Puducherry 605 014, India}
\date{\today}

\begin{abstract}
Mixed entangled states are generic resource for quantum teleportation. Optimal teleportation fidelity measures the success of quantum teleportation. The relevance of rank in the teleportation process is investigated by constructing three new maximally entangled mixed states (MEMS) of different ranks. Linear entropy, concurrence, optimal teleportation fidelity and Bell function are obtained for each of the state analytically. It is found that mixed states with higher rank are better resource for teleportation. In order to achieve a fixed value of optimal teleportation fidelity, we find that low rank states must have high concurrence. Further, for each of ranks 2, 3 and 4, we numerically generate 30000 maximally entangled mixed states. The analysis of these states reveals the existence of a rank dependent upper bound on optimal teleportation fidelity for a fixed purity. In order to achieve a fixed optimal teleportation fidelity, we find MEMS exhibit a rank dependent lower bound on concurrence. MEMS are classified in terms of their degree of nonlocality.
\end{abstract}

\maketitle

\section{Introduction}
Quantum teleportation refers to transfer of an unknown quantum state between two spatially separated observers traditionally named as Alice and Bob. Bennett {\it et. al.}~\cite{Bennett1993} proposed a protocol to teleport an unknown single qubit pure state from Alice to Bob, who share a qubit each. The two qubit state is maximally entangled pure state (Bell state) referred to as teleportation channel or a resource. Alice makes a measurement on the two qubits she has using a maximally entangled basis (Bell basis) and communicates the result of the measurement to Bob through a classical channel. Then Bob performs a local unitary operation on his qubit that completes the teleportation protocol. Quantum teleportation of an unknown qubit state has been achieved in different experiments~\cite{Bouwmeester1997,Furusawa1998,Boschi1998,Nielson1998,Kim2001,Mercikic2003}.

Preparing a pure maximally entangled state is often difficult. Due to decoherence, pure states evolve in to mixed states. When mixed entangled states are used as resource for teleportation, the optimal teleportation fidelity is no longer unity~\cite{Popescu1994}, but provides a measure for success of teleportation process. Optimal teleportation fidelity of a mixed entangled state to be used as resource for quantum teleportation should be greater than 2/3. Werner state, which is a convex sum of maximally mixed separable state and maximally entangled pure state~\cite{Werner1989} has maximum optimal teleportation fidelity for a given purity, characterized by linear entropy. Werner state is a maximally entangled mixed state (MEMS). A general class of MEMS is defined with the property that the entanglement of the state cannot be increased by a local unitary transformation~\cite{Ishizaka2000}. It is a general understanding that MEMS serve as a better resource for quantum teleportation among mixed states.  Teleportation and entanglement are nonlocal quantum phenomena. While the entanglement of mixed state used as resource is necessary for teleportation, it turns out that it is not sufficient.  For example, MJWK MEMS \cite{Munro2001} is entangled but cannot be used as a resouce for teleportation for $0 \le p \le 1/3$ (see Eq. (\ref{mjwkstate}), Eq. (\ref{mjwkf}) and fig.~\ref{fvsc}). One would expect the strength of the resource (measure of entanglement) benefit teleportation. On the contrary, we have in the MEMS category, MJWK \cite{Munro2001} state which is more entangled than the Werner state, but it's optimal telportation fidelity is less than that of Werner state for all values of linear entropy.
Further, in the category of non maximally entangled mixed states (NMEMS), we have Werner derivative\cite{Hiroshima2000} which violates Bell CHSH inequlities.  Adhikari {\it et al} constructed a state \cite{Adhikari2010} in NMEMS category which does not violate Bell CHSH inequality, but has higher fidelity compared to Werner derivative for some values of linear entropy. This leads to the fact that a mixed entangled state violating Bell CHSH inequality is sufficient for its use as resource for teleportation, but not necessary. Further, Gisin \cite{Gisin1995} obtained a bound on the fidelity  of mixed entangled states which do not admit local hidden variable description.
\begin{equation}\label{flhv}
F_{lhv}=\frac{1}{2} + \sqrt{\frac{3}{2}}\frac{{\rm arctan}(\sqrt{2})}{\pi} \approx 0.87
\end{equation}
We call this a Gisin bound and if $F > F_{lhv}$, the mixed state does not admit local hidden variable description.

In the present work, we consider maximally entangled mixed states as quantum teleportation resource. We study optimal teleportation fidelity, concurrence, linear entropy and Bell function of different MEMS and investigate their dependence on the rank.

The paper is organized as follow. Linear entropy, concurrence, optimal teleportation fidelity and Bell function are defined for mixed entangled states in section II. In section III, we present the analysis of three new maximally entangled mixed states constructed with different ranks. Here we study the optimal teleportation fidelity of each of these states as a function of linear entropy and concurrence. In order to study the role of rank of MEMS on teleportation fidelity and concurrence, we simulate 30000 MEMS of each for ranks 2, 3 and 4. Optimal teleportation fidelity of MEMS is studied with respect to its dependence on rank, concurrence and nonlocality and these results are presented in section IV. Conclusions of the work are given in section V.

\section{Definitions}
Purity of a mixed state can be measured in terms of linear entropy. For a two qubit mixed state $\rho$, the linear entropy is defined as
\begin{equation}\label{sl}
S_L = \frac{4}{3} \left[1 - Tr\left(\rho^2\right)\right]
\end{equation}
We can find whether a given two qubit state is entangled or separable using positivity under partial transpose (PPT) criterion, since PPT criterion \cite{M Horodecki1996} is both necessary and sufficient for separability of two quibt states. Further, we use concurrence as a measure of entanglement. MEMS studied in this work belong to a general class of mixed states known as X states. A general X state is given by
\begin{equation}\label{xstate}
\left(
  \begin{array}{cccc}
    \rho_{11} & 0 & 0 & \rho_{14} \\
    0 & \rho_{22} & \rho_{23} & 0 \\
    0 & \rho^{\star}_{23} & \rho_{33} & 0 \\
    \rho^{\star}_{14} & 0 & 0 & \rho_{44} \\
  \end{array}
\right)
\end{equation}
Concurrence of a general X state is given by \cite{Laura2010}
\begin{eqnarray}\label{concx}
\nonumber
C=&2& {\rm max}\left\{0, K_1, K_2\right\} \\
K_1 &=& \vert \rho_{14} \vert -\sqrt{\rho_{22}\rho_{33}}\\
\nonumber
K_2&=&\vert \rho_{23} \vert -\sqrt{\rho_{11}\rho_{44}}
\end{eqnarray}
Maximum Bell function is defined for general $X$ state as follows.
\begin{equation}
B = {\rm max}\left\{B_1, B_2\right\}
\end{equation}
where
\begin{eqnarray}
\nonumber
B_1 &=& 2\sqrt{u_1+u_2}, \ \ B_2=2\sqrt{u_1+u_3} \\
u_1&=&4 \left(\vert \rho_{14}\vert + \vert \rho_{23}  \vert \right)^2,\\
\nonumber
u_2&=&\left(\rho_{11}+\rho_{44}-\rho_{22}-\rho_{33}\right)^2,
u_3=4 \left(\vert \rho_{14}\vert - \vert \rho_{23}  \vert \right)^2
\end{eqnarray}

Following Ishizaka and Hiroshima\cite{Ishizaka2000}, we characterize a given mixed entangled state to be MEMS as follows. Let $\lambda_1$, $\lambda_2$, $\lambda_3$ and $\lambda_4$ be the eigenvalues of a given mixed entangled state $\rho$ in descending order. If the concurrence of the state is equal to $C^{\star}$, where
\begin{equation}\label{cstar}
C^{\star} = \lambda_1 - \lambda_3 - 2\sqrt{\lambda_2 \lambda_4}
\end{equation}
then the state $\rho$ is MEMS.
The optimal teleportation fidelity of an entangled mixed state is defined as~\cite{Horodecki1996}
\begin{equation}
F(\rho) =  \frac{1}{2}\left[1+\frac{N(\rho)}{3}\right]
\end{equation}
where $N(\rho)$ is defined as
\begin{equation}
N(\rho)=\sum_{i=1}^{3} \sqrt{v_i}
\end{equation}
where $v_i$ are the eigenvalues of the matrix $T^{\dagger}T$ and the elements of correlation matrix $T$ are given by
\begin{equation}
T_{nm} = Tr\left(\rho \sigma_n \otimes \sigma_m\right)
\end{equation}

\section{Maximally Entangled Mixed States}
We consider Werner state as reference state because it is a construction involving the maximally entangled pure state and maximally mixed separable state. In addition, the optimal telportation fidelity of Werner state is maximum for a given value of linear entropy, compared to all the mixed entangled states considered as a resource for teleportation in the literature. Our aim is to understand the role played by mixing (characterized by linear entropy), entanglement (measured in terms of concurrence), rank and Bell function on the optimal teleportation fidelity. For this purpose, we construct three new mixed entangled states as variation of Werner state.

Werner state, a rank 4 MEMS is given by
\begin{equation}\label{werner}
\rho_W = (1-p)\frac{I_4}{4} + p \vert \psi^{+} \rangle \langle \psi^{+} \vert
\end{equation}
where $\vert \psi^{+} \rangle$ is maximally entangled pure state given in terms of the computational basis by
\begin{equation}
\vert \psi^{+} \rangle = \frac{1}{\sqrt{2}}\left[\vert 01 \rangle + \vert 10 \rangle  \right]
\end{equation}
It is known that the Werner state is entangled for $1/3 < p \le 1$ and violates Bell CHSH inequality for $1/\sqrt{2} < p \le 1$, i.e., its bell function $B(p)=2\sqrt{2}p$. The linear entropy of Werner state is $S_L(p)=1-p^2$, concurrence is $C(p)=(3p-1)/2$ and optimal teleportation fidelity is $F(p)=(p+1)/2$.

The other maximally entangled mixed state known is MJWK state given by~\cite{Munro2001}
\begin{equation}\label{mjwkstate}
\left(
  \begin{array}{cccc}
    \gamma(p) & 0 & 0 & \frac{p}{2} \\
    0 & 1-2\gamma(p) & 0 & 0 \\
    0 & 0 & 0 & 0 \\
    \frac{p}{2} & 0 & 0 & \gamma(p) \\
  \end{array}
\right)
\end{equation}
where $\gamma(p)=1/3$ for $0 \le p < 2/3$ and $\gamma(p)=p/2$ for $2/3 \le p \le 1$. The linear entropy of this state is given by
\begin{equation}\label{mjwksl}
S_L(p)=\left\{
  \begin{array}{ll}
    (8-6p^2)/9, & {0 \le p < 2/3;} \\
    (8p-8p^2)/3, & {2/3 \le p \le 1.}
  \end{array}
\right.
\end{equation}
MJWK is a rank 3 state for $0 \le p < 2/3$ and a rank 2 state for $2/3 \le p < 1$. The concurrence of MJWK state is $C(p)=p$. The optimal teleportation fidelity is given by
\begin{equation}\label{mjwkf}
F(p)=\left\{
  \begin{array}{ll}
    (5+3p)/9, & {0 \le p < 2/3;} \\
    (2p+1)/3, & {2/3 \le p \le 1.}
  \end{array}
\right.
\end{equation}
It can be observed that MJWK state is entangled for all $0 \le p \le 1$, but it can be used as a resource for quantum teleportation only in the range $1/3 < p \le 1$. Concurrence of MJWK~\cite{Munro2001} state is greater than Werner state for a given value of linear entropy. On the other hand the optimal teleportation fidelity of Werner state is greater than MJWK state for all values of linear entropy. This implies that the entanglement is not directly aiding the teleportation process in the case of mixed entangled states.

Since the ranks of Werner state and MJWK states are different, we investigate the role of rank of MEMS on the teleportation process. We construct three new mixed entangled states. For that purpose, we obtain a MEMS by taking a partial trace~\cite{M Nielsen} of a three qubit W-state~\cite{A Zeilinger} given by
\begin{equation}\label{ptwstate}
\rho_M = Tr_1\left[\frac{1}{\sqrt{3}}\left(\vert 100 \rangle + \vert 010 \rangle + \vert 001\rangle \right)\right]
\end{equation}
We construct our fist mixed entangled state by replacing the maximally mixed pure state of Werner state by the above MEMS.
\begin{equation}\label{first}
\rho_1= (1-p) \rho_M + p \vert \psi^{+} \rangle \langle \psi^{+} \vert
\end{equation}
Using PPT criterion, we show tha the state $\rho_1$ is entangled for $0 \le p \le 1$. The rank of $\rho_1$ is 2. We find that it is MEMS for $0 \le p \le 1$.

Our second state involves replacing maximally mixed separable state in Werner state with non maximally mixed separable state and maximally entangled pure state with MEMS. The non maximally mixed separable state is obtained by taking a partial trace over a three qubit state that belongs to GHZ~\cite{Greenberger} family of states.
\begin{equation}\label{ptghz}
\rho_G = Tr_2\left(\frac{1}{\sqrt{2}}\left[\vert 100 \rangle + \vert 011 \rangle \right]\right)
\end{equation}
and the second state is given by
\begin{equation}\label{third}
\rho_2=(1-p)\rho_G + p\rho_M
\end{equation}
Our second state is similar in construction to the NMEMS state constructed by Adhikari {\it et. al.}~\cite{Adhikari2010}, but the state is different since $\rho_G$ is different. Adhikari {\it et. al} constructed $\rho_G$ by taking a partial trace over a three qubit GHZ state $(\vert 000 \rangle +\vert 111 \rangle)/\sqrt{2}$. However, if we take partial trace over first qubit in Eq.(\ref{ptghz}), it will be same as the one constructed in~\cite{Adhikari2010}.

The rank of $\rho_2$ is 3. We find $\rho_2$ is entangled for $0 < p \le 1$, but is MEMS only in the range $3/5 \le p \le 1$. We restrict our studies to this range where the state is MEMS.
We construct our third state by replacing the maximally entangled pure state in Werner state by a  MEMS $\rho_M$ given in Eq.(\ref{ptwstate}).
\begin{equation}\label{second}
\rho_3 = (1-p)\frac{I_4}{4} + p \rho_M
\end{equation}
The rank of $\rho_3$ is 4. The state $\rho_3$ is entangled for $3/(1+2\sqrt{5}) < p \le 1$ and is also MEMS in this region.
Linear entropy, concurrence and optimal teleportation fidelity of three new states are obtained analytically and the expressions are given in table~I.
\begin{table}
\caption{Analytical Expressions for linear entropy, concurrence and optimal teleportation fidelity of three new MEMS}
\begin{center}
\begin{tabular}{|l|l|l|l|l|}
\hline
 State & Rank & $S_L(p)$  & $C(p)$ & $F(p)$ \\
\hline
$\rho_1$ & 2 &  $\frac{8}{27}(1-p)(p+2)$ & $\frac{2+p}{3}$ & $\frac{4p+14}{18}$ \\
$\rho_2$ & 3 & $\frac{2}{3}\left(1+\frac{2p}{3}-\frac{7p^2}{9}\right)$ & $\frac{2p}{3}$ & $\frac{2p+12}{18}$ \\
$\rho_3$ & 4 & $1 - \frac{11p^2}{27}$ & $\frac{2p}{3}- \sqrt{\frac{(1-p)(3+p)}{12}}$ & $\frac{9+5p}{18}$ \\
\hline
\end{tabular}
\end{center}
\end{table}

\begin{figure}
\centering
\includegraphics[width=3.5in]{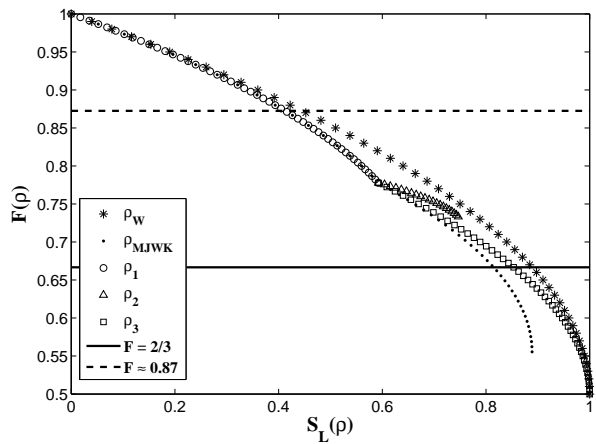}
\caption{Variation of optimal teleportation fidelity with linear entropy for Werner state, MJWK state and three new MEMS given in table~I}\label{fvssl}
\end{figure}

By eliminating the parameter $p$, we present the variation of optimal teleportation fidelity with linear entropy for three new states, Werner state and MJWK state in fig.~\ref{fvssl}. We observe that optimal teleportation fidelity is highest for Werner state for all values of linear entropy. We observe from fig.~\ref{fvssl} that the optimal teleportation fidelity of the state $\rho_1$ coincides with that of MJWK state. However the state $\rho_1$ can be a resource for teleportation in the linear entropy range $0 \le S_L \le 16/27$. The state $\rho_2$ is defined in the linear entropy range $16/27 < S_L \le 1$. Optimal teleportation fidelity of this state is close to MJWK state when linear entropy is close to 16/27 and increases from MJWK state for higher values of $S_L$ and gets close to Werner state as $S_L$ approaches unity. The state $\rho_3$ has higher fidelity compared to $\rho_2$ for some range of linear entropy. Our analysis indicates that for high value of linear entropy, high rank state is a better resource for teleportation.

\begin{figure}
\centering
\includegraphics[width=3.5in]{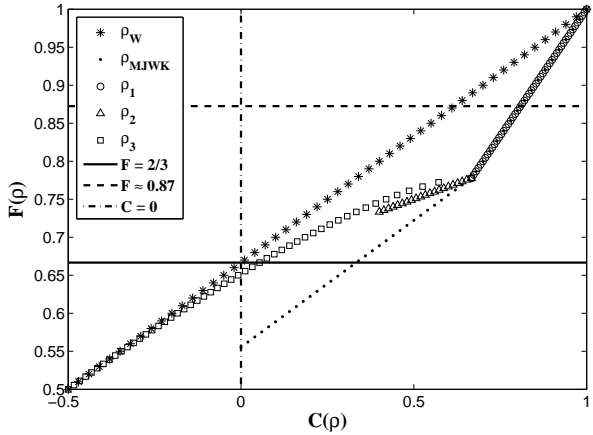}
\caption{Optimal teleportation fidelity as a function of concurrence of Werner state, MJWK state and three new MEMS given in table~I}\label{fvsc}
\end{figure}

Optimal teleportation fidelity of three new states together with Wener state and MJWK state as a function of concurrence is shown in fig.~\ref{fvsc}. We observe that in order to achieve a fixed value of optimal teleportation fidelity, low rank state must have higher concurrence than a state of high rank. In order to understand the dependence of the optimal teleportation fidelity on the rank of MEMS, we numerically generate MEMS of different ranks and study their optimal teleportation fidelities as a function of linear entropy and concurrence.

\section{MEMS resource for teleportation: Rank dependence}
Ishizaka and Hiroshima~\cite{Ishizaka2000} proposed a class of MEMS. We generated 30000 random density matrices corresponding to each of the ranks 2, 3 and 4. For each MEMS, we computed linear entropy, concurrence, optimal teleportation fidelity and Bell function.
\begin{figure}
\centering
\includegraphics[width=3.5in]{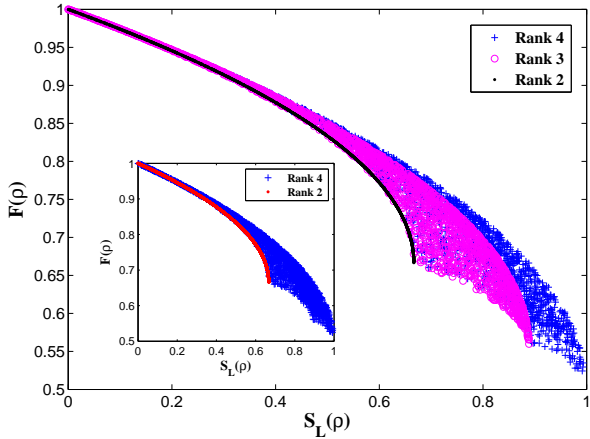}
\caption{Optimal teleportation fidelity as a function of linear entropy for randomly generated MEMS of rank 2, 3 and 4. 30000 MEMS of each rank are randomly generated. Only 1000 of them for each rank are shown here for presentation clarity. Inset shows the same figure without MEMS of rank 3}\label{ranksl}
\end{figure}

\begin{figure}
\centering
\includegraphics[width=3.5in]{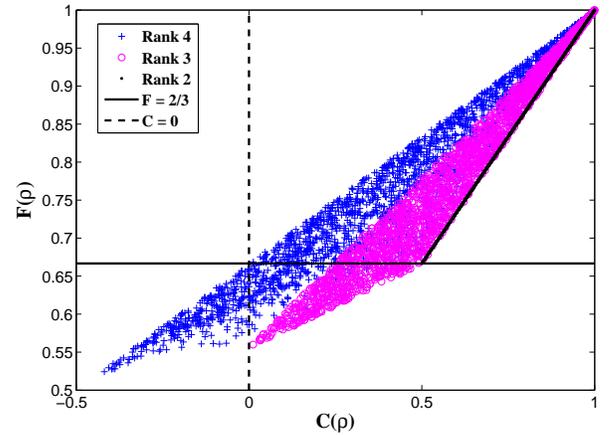}
\caption{Optimal teleportation fidelity as a function of concurrence for randomly generated MEMS of rank 2, 3 and 4. 30000 MEMS of each rank are randomly generated. Only 1000 of them for each rank are shown here for presentation clarity.}\label{rankc}
\end{figure}

In fig~\ref{ranksl}, we present the optimal teleportation fidelity as a function of linear entropy of 1000 MEMS for each of rank. We present only 1000 MEMS for each rank for clarity in presentation and they cover representative regions corresponding to each rank in Optimal teleportation fidelity - linear entropy plane. We observe that for a given value of linear entropy, optimal teleportation fidelity is maximum for MEMS of rank 4. Werner state is of rank 4 and it exhibits highest optimal telportation fidelity for any value of linear entropy can be understood based on this observation. It should be noted that in the $F-S_L$ plane, MEMS of rank 4 overlap the regions of MEMS of lower ranks. This can be seen from the inset of fig~\ref{ranksl} where only MEMS of rank 2 and rank 4 are shown.  Our numerical results indicate that for a given value of linear entropy, there exists an upper bound on the optimal teleporation fidelity for a MEMS of a given rank. This upper bound increases with increase in the rank of MEMS.

Optimal teleportation fidelity of randomly generated MEMS of different ranks as a function of concurrence is shown in fig.~\ref{rankc}. It can be observed that among randomly generated states, states of rank 2 and 3 are always entangled with positive value of concurrence. All MEMS of rank 2 qualify to be resource for quantum teleportation with optimal teleportation fidelity greater than 2/3. Our numerical results reveal that, in order to achieve a given value of optimal teleportation fidelity, concurrence should be higher for lower rank MEMS. In other words, there exists a rank dependent lower bound on the concurrence of MEMS for a fixed value of optimal teleportation fidelity. The lower bound decreases with increase in rank. Based on these results, we can understand the lower optimal teleportation fidelity of MJWK state in spite of its concurrence being higher compared to Werener state.

\begin{figure}
\centering
\includegraphics[width=3.5in]{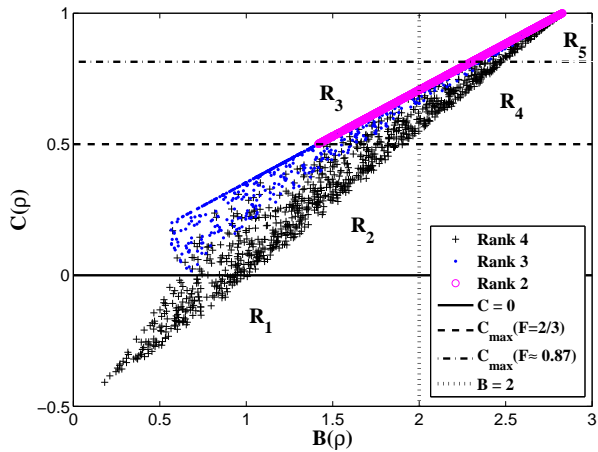}
\caption{Concurrence as a function of Bell function for randomly generated MEMS of rank 2, 3 and 4. 30000 MEMS of each rank are randomly generated. Only 1000 of them for each rank are shown for presentation clarity. $R_1$: Mixed states that are not entangled; $R_2$: MEMS that cannot be used for quantum teleportation; $R_3$: MEMS that can be used as resource for quantum teleportation, but do not violate Bell CHSH inequality; $R_4$: MEMS that obey Bell CHSH inequality, but optimal teleportation fidelity is less than Gisin bound; $R_5$: MEMS whose optimal teleportation fidelity is greater than Gisin bound.}\label{cvsb}
\end{figure}

Figure~\ref{cvsb} presents concurrence versus Bell function. We find that the minimum Bell function for a given value of concurrence corresponds to lower rank MEMS, consistent with what is reported in \cite{Verstraete}. Further we characterize MEMS with respect to the degree of nonlocality. In $C-B$ plane, states that lie below $C=0$ line are not entangled (region $R_1$). In the figure $C_{max}(F=2/3)$ is maximum concurrence of MEMS of all ranks corresponding to optimal teleportation fidelity $F=2/3$. States above $C=0$ line and below $C=C_{max}(F=2/3)$ line are entangled, but cannot be used as resource for teleportation since their optimal teleportation fidelity is less than 2/3. This region is denoted as $R_2$. States above $C=C_{max}(F=2/3)$ and left of $B=2$ vertical line correspond to the ones that can be used as resource for quantum teleportation, but do not violate Bell CHSH inequality (region $R_3$). $C_{max}(F \approx 0.87)$ refers to maximum concurrence of all MEMS for which optimal teleportation fidelity is given by Eq.(\ref{flhv}). To the right of $B=2$ vertical line and below $C=C_{max}(F \approx 0.87)$ line lie the states that violate Bell CHSH inequality, but fidelity is less than Gisin bound (region 4) and states above $C=C_{max}(F \approx 0.87)$ line correspond to strongly nonlocal states that do not admit local hidden variable description. This suggests an ordering in terms of the degree of nonlocality of mixed states. Entangled states , states that can be used as teleportation resource, states that violate Bell CHSH inequality and states that do not admit local hidden variable description can be regarded as the order in which degree of nonlocality is increasing.

\section{Conclusions}
We systematically studied the relevance of rank of a two qubit mixed entangled states as a resource for quantum teleportation. We constructed three new maximally entangled mixed states of different ranks and obtained analytical expressions for linear entropy, concurrence and optimal teleportation fidelity. The results are compared with two well known MEMS, viz., Werner state and MJWK state. It is found that for any given purity of MEMS, maximum optimal teleportation fidelity can be achieved by MEMS of rank 4. We randomly generated MEMS of different ranks and investigated their optimal teleportation fidelity as a function of linear entropy and concurrence. Our numerical investigations on MEMS of different ranks revealed the existence of an upper bound on optimal teleportation fidelity for MEMS of a given rank and fixed linear entropy. This upper bound increases with increase in rank of the state. Further, in order to achieve a fixed value of optimal teleportation fidelity, there exists a lower bound on the concurrence for MEMS of a given rank. This lower bound decreases with increase in the rank. Entangled states, states that can be used as teleportation resource, states that violate Bell CHSH inequality and states that have optimal teleportation fidelity greater than Gisin bound is the order in which degree of nonlocality increases.

\begin{center}
{\bf ACKNOWLEDGEMENTS}
\end{center}

We acknowledge Alok Sharan for useful discussions.

\end{document}